\documentclass[graybox]{svmult}

\usepackage{mathptmx}       
\usepackage{helvet}         
\usepackage{courier}        
\usepackage{type1cm}        
\usepackage{makeidx}         
\usepackage{graphicx}        
\usepackage{multicol}        
\usepackage[bottom]{footmisc}

\usepackage{bm}

\makeindex             

\newcommand{\tb}{Budav\'ari}
\newcommand{\like}{\mathcal{L}}

\newcommand{\drxn}{\omega}  
\newcommand{\tlabel}{\lambda}  

\begin{document}

\title*{Commentary on Bayesian coincidence assessment (cross-matching)}
\author{Thomas J. Loredo}
\institute{Thomas J. Loredo \at Center for Radiophysics \& Space Research,
Cornell University, Ithaca, NY 14853-6801, \email{loredo@astro.cornell.edu}}
%
%
\maketitle

\abstract*{This contribution is a commentary on Tam\'as Budav\'ari's paper,
``On statistical cross-identification in astronomy,'' presented at the {\em
Statistical Challenges in Modern Astronomy V} conference held at
Pennsylvania State University in June 2011.}

\begin{quote}\em
This paper is an invited commentary on Tam\'as Budav\'ari's presentation, ``On
statistical cross-identification in astronomy,'' for the {\em Statistical
Challenges in Modern Astronomy V} conference held at Pennsylvania State
University in June 2011.  I begin with a brief review of previous work on
probabilistic (Bayesian) assessment of directional and spatio-temporal
coincidences in astronomy (e.g., cross-matching or cross-identification of
objects across multiple catalogs).  Then I discuss an open issue in the recent
innovative work of Budavari and his colleagues on large-scale probabilistic
cross-identification:  how to assign prior probabilities that play an
important role in the analysis.  With a simple toy problem, I show how
Bayesian multilevel modeling (hierarchical Bayes) provides a principled
framework that justifies and generalizes pragmatic rules of thumb that have
been successfully used by Budavari's team to assign priors.
\end{quote}

\noindent
\tb's paper reviews the key concepts of a recent body of research by him and
his colleagues on Bayesian cross-matching of astronomical object catalogs.
When object directions have quantified uncertainties (e.g., error circles
with confidence levels), this approach offers significant advantages over
more conventional approaches that attempt to assess directional coincidences
using ad hoc statistics (e.g., nearest neighbor angles, counts in cones,
$\chi^2$-based statistics, likelihood ratios) and $p$-values.  In the
mid-1990s gamma-ray burst (GRB) astronomers developed essentially the same
approach for assessing evidence for repetition of GRBs
\cite{LLW96,GL96,GLQ98}, and for association of GRBs with unusual supernovae
\cite{GLM99}.  This work pre-dates the discovery of GRB X-ray afterglows;
the available GRB data provided direction estimates with large uncertainties
($5^\circ$ to $25^\circ$ error circles for directions from BATSE data;
many-arc-minute error boxes for interplanetary network direction estimates).
\tb\ seeks to cross-match optical and UV catalogs that are much larger in
size than GRB catalogs, and with much more accurate directions.  This is a
complementary regime, raising unique challenges for Bayesian
cross-matching---especially computational challenges---that \tb's team is
addressing with innovative techniques just briefly touched on in his paper
(e.g., see \cite{BS08,Buda+09,Buda11,Kere+10}).

For this commentary I am taking a cue from \tb's Discussion section, where
he states that a ``fully Bayesian solution would include a hierarchical
model'' but that ``the computational requirements do not justify the extra
work.''  Luo, Loredo \& Wasserman (\cite{LLW96}, LLW96) developed a
hierarchical (multilevel) Bayesian framework for assessing directional and
temporal coincidences in a GRB catalog; an exact calculation was indeed
impossible, and LLW96 had to rely on an unsatisfying approximate treatment. 
However, for an important issue that \tb\ discusses, a simple
multilevel model is both computationally accessible, and illuminating.

Specifically, \tb\ highlights the important role of the prior probability
for association, $P_0$, in Bayesian cross-matching; it is needed to convert
marginal likelihoods (or Bayes factors) for association to posterior
probabilities (or odds).  But \tb's $P_0$ is determined using the data; it
cannot really be a prior probability.  A multilevel model not only enables
estimation of $P_0$, but also can account for its uncertainty, which should
play a role in assessing the method's performance in simulation studies
(e.g., determining whether the discrepancy between estimated and true
association fractions in \tb's Fig.~4 is acceptable).  A multilevel
treatment also illuminates other issues important for probabilistic
cross-matching.  In the limited space available here I describe a simple
example calculation illustrating the main idea; a more complete and general
treatment will be published elsewhere.

Suppose we have a ``target'' catalog of $N_t$ newly detected objects,
and we would like to determine if some or all of them are
associated with any of $N_c$ previously detected objects in a candidate host
or counterpart catalog spanning the same region of the sky.
From the target observations, analysis of the data associated with object
number $i$ produces a likelihood function, $\ell_i(\drxn)$, for its
direction, $\drxn$; the target catalog provides summaries of these
likeihood functions (e.g., best-fit directions and error circles
when uncertainties can be accurately described by Fisher distributions). 
Similarly, $m_k(\drxn)$ is the likelihood function for the direction to
object $k$ in the candidate host catalog. Suppose the cataloged hosts are a
sample from a large population with a known (or well-estimated) directional
distribution, $\rho_c(\drxn)$, e.g., an isotropic distribution with $\rho_c
= 1/4\pi$.  Then the posterior distribution for the direction to candidate
host object $k$ is $\rho_c(\drxn) m_k(\drxn)/Z_k$, where the
normalization constant $Z_k = \int d\drxn\,\rho_c(\drxn) m_k(\drxn)$.  The
marginal likelihood that target $i$ is associated with host $k$ (thus
sharing a common direction) is
\begin{equation}
h_{ik} = 
    \int d\drxn \frac{\rho_c(\drxn) m_k(\drxn)}{Z_k} \ell_i(\drxn).
\label{mlike-ik}
\end{equation}
The marginal likelihood that target $i$ is instead from a background
population of hosts with direction distribution $\rho_0(\drxn)$ is
\begin{equation}
g_i = \int d\drxn\,\rho_0(\drxn) \ell_i(\drxn).
\label{mlike-back}
\end{equation}
The Bayes factor in favor of association of target $i$ with host $k$
versus a background source is $b_{ik} = h_{ik}/g_i$.  When the direction
likelihoods are proportional to Fisher distributions and the host
and background densities are isotropic, this corresponds to \tb's $B$ (also
derived earlier by LLW96 and \cite{GL96}).

Of course, we do not know a priori which candidate host to assign to
each target.  The marginal likelihood that target $i$ is associated
with {\em one} of the candidate hosts must account for this uncertainty
by introducing a prior probability for the host choice, say $1/N_c$, and
marginalizing over $k$; the resulting marginal likelihood is
\begin{equation}
h_i
 = \frac{1}{N_c} \sum_k h_{ik}
 = \frac{1}{N_c} \sum_k
    \int d\drxn \frac{\rho_c(\drxn) m_k(\drxn)}{Z_k} \ell_i(\drxn).
\label{mlike-assoc}
\end{equation}

\tb\ introduced a prior probability for association, $P_0$, in order to
convert marginal likelihoods (or Bayes factors) to posterior probabilities
(or odds).  Using intuitively appealing arguments, he develops equations to
determine a value for $P_0$, but they use the data, and thus $P_0$ is not
really a prior probability, and his posterior probabilities are not 
formally valid.  To better motivate and extend \tb's appealing
results, we make the association model a multilevel model, introducing a
population parameter that we will estimate from the data.

Define the target population association parameter, $\alpha$, as the
probability that a randomly selected target comes from the population of
cataloged candidate hosts (so $1-\alpha$ is the probability that a target
comes from the background).  Were $\alpha$ known, the posterior probability
that target $i$ is associated with one of the hosts would be
\begin{equation}
p_i(\alpha) = \frac{\alpha h_i}{(1-\alpha) g_i + \alpha h_i}.
\label{prob-assoc}
\end{equation}
But typically $\alpha$ will {\em not} be known a priori; in fact, estimating
$\alpha$ may be a significant scientific goal.  The likelihood function
for $\alpha$ is the probability for the target data, given $\alpha$ and the
host catalog information; using the above results, it is
\begin{equation}
\like(\alpha) = \prod_{i=1}^{N_t} \left[
  (1-\alpha) g_i + \alpha h_i \right].
\label{like-alpha}
\end{equation}
A straightforward calculation shows that the maximum-likelihood value
of $\alpha$, $\hat\alpha$, satisfies the following equation:
\begin{equation}
\sum_i p_i(\hat\alpha) = \hat\alpha N_t.
\label{alpha-hat}
\end{equation}
This is an intuitively appealing result:  for the maximum-likelihood value
of $\alpha$, the sum of the association probabilities is equal to the
expected number of targets with associations.

To see the connection with \tb's rule for assigning $P_0$ (his
equation~(10)), suppose the data provide direction estimates with
uncertainties that are small compared with the angles between hosts.  Then
the sum in the marginal likelihood for association for a target object,
equation~(\ref{mlike-assoc}), will typically be dominated by just one term, so
\begin{equation}
h_i \approx \frac{1}{N_c} 
    \int d\drxn \frac{\rho_c(\drxn)m_{k(i)}(\drxn)}{Z_{k(i)}} \ell_i(\drxn),
\label{mlike-host-1}
\end{equation}
where $k(i)$ specifies the index of the host that is the nearest neighbor to
target $i$ (in the sense of having the largest marginal likelihood term).
If we use this approximation for $h_i$ in equation~(\ref{prob-assoc}) for
$p_i(\alpha)$, then equation~(\ref{alpha-hat}) becomes equivalent to
\tb's equation~(10) (identifying $\hat\alpha$ with his $P_0$, $p_i$ with his
$P$, and $\hat\alpha N_t$ with his $N_\star$), for the case of two catalogs.

This calculation does more than simply justify \tb's intuitive arguments
for setting $P_0$.  One concrete benefit is that it enables accounting
for uncertainty in $\alpha$.  Combined with a prior for $\alpha$, the
likelihood function in equation~(\ref{like-alpha}) produces a posterior
for $\alpha$.  If the prior is not highly informative, the posterior
will be asymptotically normal, with a mean close to $\hat\alpha$ and
a variance, $\sigma_\alpha^2$, that can be found by calculating the second
derivative of $\ln[\like(\alpha)]$ at $\hat\alpha$; the result is
\begin{equation}
\frac{1}{\sigma_\alpha^2}
  \;=\; \frac{1}{\hat\alpha(1-\hat\alpha)} \sum_i \left(\hat p_i - \hat\alpha\right)^2
  \;=\; \frac{N_t}{\hat\alpha(1-\hat\alpha)}
    \left[ \frac{1}{N_t}\sum_i \hat p_i^2 - \left(\frac{\sum_i \hat p_i}{N_t}\right)^2
    \right],
\label{alpha-var}
\end{equation}
where $\hat p_i \equiv p_i(\hat\alpha)$.  Two limiting cases are
illuminating.  Suppose first that the target positions have very large
uncertainties.  In the limit where $\ell_i(\drxn) \rightarrow C$, a
constant (i.e., uninformative data), we have $g_i = h_i = C$.  The Bayes factor for association of each
object is unity (indicating the data provide no information to alter prior
probabilities), and the likelihood function for $\alpha$ is flat, so there
is no unique $\hat\alpha$ value.  The right hand side of equation~(\ref{alpha-var})
vanishes, implying divergence of the variance (actually, the asymptotic
approximation is not valid with a flat likelihood function). 
The data provide no information about the association fraction in this case,
as one would expect.  Now consider the opposite limit where the direction
uncertainties are small, leading to unambiguous associations (very large
Bayes factors), so that for values of $\alpha$ away from zero or unity,
$p_i \approx 0$ or $1$.  In this case, equation~(\ref{alpha-hat}) tells us
that $\hat\alpha = N_+/N_t$, where $N_+$ is the number of targets with
$p_i \approx 1$.  Equation~(\ref{alpha-var}) indicates that in this limit,
$\sigma_\alpha \rightarrow 1/\sqrt{N_t}$, again an intuitively reasonable
result.  For intermediate cases, where there is evidence for
associations but with some ambiguity, the uncertainty in $\alpha$ will be
larger than ``root-$N$,'' by an amount depending on the variance between the
$\hat p_i$ values and $\hat\alpha$.  Calculating $\sigma_\alpha$ for the
SDSS--Galex example in \tb's Section~4 may be helpful in assessing the
discrepancy between the estimated and input values of $P_0$.

In the SDSS--Galex example, $P_0$ was over-estimated; \tb\ attributes this
to confusion due to chance proximity of objects in each catalog.  But one
of the aims of probabilistic modeling of directional coincidences is to
account for this sort of confusion.  An accurate Bayesian calculation will
account for it, resulting in no significant bias in estimation of the
association fraction, but possibly increased $\alpha$ uncertainty when the
directional uncertainties lead to significant counterpart confusion.  When a
particular target has multiple plausible associations, the probability for
association will be split across them.
One way to see how the Bayesian calculation handles counterpart ambiguity is
to rewrite the likelihood function to more explicitly display how it
accounts for each possible association.  First introduce unifying notation
for the components in the likelihood factor for a particular target: define
weights $w_k$, with $w_0 = 1-\alpha$ and $w_k = \alpha/N_c$ for $k=1$ to
$N_c$, and let $h_{ik} = g_i$ when $k=0$.  Also introduce target labels
$\tlabel_i$ that take values from $0$ to $N_c$.  Then
equation~(\ref{like-alpha}) can be written as
\begin{equation}
\like(\alpha) 
  \;=\; \prod_{i=1}^{N_t} 
      \sum_{\tlabel_i=0}^{N_c} w_{\tlabel_i} h_{i\tlabel_i}
  \;=\; \sum_{\tlabel_1\ldots \tlabel_{N_t}}
      \left(\prod_k w_k^{m_k(\tlabel)}\right)
      \prod_i h_{i\tlabel_i},
\label{like-sum-prod}
\end{equation}
where the last sum is over all label assignments, and $m_k(\tlabel)$ is the
multiplicity for host $k$, counting the number of targets with $\tlabel_i=k$
in a particular term of the sum.  This sum-of-products decomposition
displays the likelihood as a weighted sum of terms considering every
possible assignment of targets to candidate hosts. If we adopt the
best-candidate approximation of equation~(\ref{mlike-host-1}), only a small
fraction of the terms is considered; when confusion is important, additional
terms in $h_i$ should be kept so that the calculation accounts for all
plausible associations.

Equation~(\ref{like-sum-prod}) also reveals an unsatisfactory aspect of the
model I have described here:  it allows for all possible host multiplicities,
in particular, it allows for assigning two targets to the same host.  In
some settings this is desirable, e.g., for constraining GRB
repetition, or for determining whether ultra-high energy cosmic rays come
from nearby active galaxies.  But in many settings---including the
SDSS--Galex case---it is only meaningful to assign targets to distinct hosts. 
This argues that the sum-of-products version of the likelihood function is
the more fundamental representation to use for building coincidence
assessment models; for the SDSS--Galex case, the sum over labels would be
constrained to ensure distinct associations.  This is why LLW96 adopted this
representation for developing a general framework for spatio-temporal
coincidence assessment.

As a final remark on the value of an explicit multilevel model for
associations, recall that we needed to assign a prior probability for the
host choice, taken as $1/N_c$ in equation~(\ref{mlike-assoc}); in the
sum-of-products version of the likelihood function, this assignment appears
in the $w_k$ factors.  More generally, the candidate host prior may not be
constant; it could depend, for example, on host distances and luminosities,
and this affects estimation of $\alpha$.  It is straightforward to account
for this in a multilevel model, though it can complicate the calculations. 
The paper in these proceedings by Soiaporn et al.\ briefly describes work by
my team based on just such a model, developed to assess evidence for
association of ultra-high energy cosmic rays with local active galaxies.

\tb\ developed his Bayesian approach from scratch, unaware of earlier work
on the problem in the GRB literature.  In fact, that work was well-hidden,
tersely presented in short papers in conference proceedings.  More extensive
treatments did not follow because it proved extremely difficult to get
funding to further develop the approach; reviewers expressed strong
skepticism of Bayesian methods.  To cite one ironically relevant example,
the report from a 2005 NVO proposal review panel asserted that the Bayesian
approach offered ``nothing new'' for the problem, and that its
implementation ``would not be much more than a `few-liner' addition to {\tt
Xmatch},'' the $\chi^2$-based NVO cross-match algorithm now made obsolete by
\tb's Bayesian algorithm. With this frustrating history, it has been a
delight to see \tb's team not only rediscover the approach, but also make
significant and highly nontrivial statistical and computational innovations
mating it to the needs of VAO users.

\begin{acknowledgement}
Ira Wasserman helped me develop a framework for Bayesian
coincidence assessment in the mid 1990s; that work was partially supported
by NASA grant NAG~5-2762.  Shan Luo helped us with early calculations. 
Currently, David Chernoff and statisticians David Ruppert and Kunlaya
Soiaporn are helping us take the approach much further; our work together is
funded by an interdisciplinary NSF grant, AST-0908439.  I am grateful to all
of these collaborators and funding agencies for their contributions to the
research informing this commentary.
\end{acknowledgement}
%


\newcommand{\apj}{Astrophysical Journal\ }
\newcommand{\aaps}{Astronomy and Astrophysics, Supplement\ }



\end{document}